\documentclass{osa-article}
\journal{optica}
\articletype{Research Article}

\usepackage{graphicx}
\usepackage{amsmath}
\usepackage{amssymb}
\usepackage{amsfonts}
\usepackage{textcomp}

\usepackage{chngcntr}

\newcommand{\Er}{\ensuremath{{\cal E}_{\rm R}}}
\newcommand{\Eo}{\ensuremath{{\cal E}_{1,2,3}}}

\begin{document}

\title{Rabi oscillations of a quantum dot exciton coupled to acoustic phonons: coherence and population readout}

\author{Daniel Wigger\authormark{1,*}, Christian Schneider\authormark{2}, Stefan Gerhardt\authormark{2}, Martin Kamp\authormark{2}, Sven H\"ofling\authormark{2,3}, Tilmann Kuhn\authormark{1}, and Jacek~Kasprzak\authormark{4,*}}

\address{
\authormark{1} Institut f\"{u}r Festk\"{o}rpertheorie, Universit\"{a}t M\"{u}nster, 48149 M\"{u}nster, Germany\\
\authormark{2} Technische Physik, Universit\"{a}t W\"{u}rzburg, 97074 W\"{u}rzburg, Germany\\
\authormark{3} SUPA, School of Physics and Astronomy, University of St. Andrews, St. Andrews KY16 9SS, UK\\
\authormark{4} Univ. Grenoble Alpes, CNRS, Grenoble INP, Institut N\'{e}el, 38000 Grenoble, France
}

\email{\authormark{*}d.wigger@wwu.de, jacek.kasprzak@neel.cnrs.fr}


\begin{abstract}
While the advanced coherent control of qubits is now routinely carried out in low frequency (GHz) systems like single spins, it is far more challenging to achieve for two-level systems in the optical domain. This is because the latter evolve typically in the THz range, calling for tools of ultrafast, coherent, nonlinear optics. Using four-wave mixing micro-spectroscopy, we here measure the optically driven dynamics of a single exciton quantum state confined in a semiconductor quantum dot. In a combined experimental and theoretical approach, we reveal the intrinsic Rabi oscillation dynamics by monitoring both central exciton quantities, i.e., its occupation and the microscopic coherence, as resolved by the four-wave mixing technique. In the frequency domain this oscillation generates the Autler-Townes splitting of the light-exciton dressed states, directly seen in the four-wave mixing spectra. We further demonstrate that the coupling to acoustic phonons strongly influences the FWM dynamics on the picosecond timescale, because it leads to transitions between the dressed states.
\end{abstract}


\section{INTRODUCTION}
The coherent control of individual spins and excitons in quantum
dots (QDs) has been the scope of semiconductor quantum optics since
many years. To unveil the potential lying in QDs and other single
photon emitters for high-speed optical quantum technology, it is
urgent to bring the manipulation schemes to perfection. This is
nowadays approached using individual spins in electrically
defined~\cite{Mi2018aco} and
epitaxial~\cite{Press2008com,DeGreve2011ult} QDs, as well as in
color centers in diamond~\cite{Gao2015coh,Zhou2017coh}, which is
conditioned by setting up robust spin-photon interfacing. This
difficulty can be mitigated using QD excitons: because of their
direct coupling to light, the related coherent control protocols
operate on the picosecond (ps) timescale~\cite{Fras2016mul}. The 
context of ultrafast control points to the subject of our
work, focusing on optically driven coherent dynamics of the exciton
state efficiently coupled to longitudinal acoustic (LA) phonons of
the GaAs host matrix. Previous research has shown that by observing
Ramsey fringes~\cite{bonadeo1998coh,stufler2006ram} the quantum
state of single localized excitons can be navigated on the Bloch
sphere by varying the delay between two laser pulses with ps
duration. It has also been demonstrated that by increasing the
applied laser power the final exciton occupation after a single
pulse can be adjusted~\cite{stievater2001rab,ramsay2010dam} by
performing Rabi rotations. Excitations with chirped laser pulses have been used to
realize robust state preparations via the so called rapid adiabatic
passage
effect~\cite{wu2011pop,simon2011rob,ardelt2014dis,quilter2015pho,bounouar2015pho,kaldewey2017coh,kaldewey2017dem}.
A pump-probe experiment~\cite{boyle2009bea} and time-resolved
resonance fluorescence~\cite{schaibley2013dir} have been used to
study the optically driven Rabi oscillation dynamics of the QD
exciton occupation. Rabi oscillations have also been
investigated in larger systems, i.e., in QD
ensembles~\cite{borri2000tim,kolarczik2013qua,karni2013rab}, quantum
wells~\cite{cundiff1994rab,choi2010ult} and quantum dash
ensembles~\cite{capua2014rab}. Note, that these Rabi oscillations
have to be differentiated from Rabi
rotations~\cite{stievater2001rab,forstner2003pho,patton2005coh,ramsay2010dam,mccutcheon2010qua,
wigger2017exp,poltavtsev2017dam,suzuki2018det}. The oscillations
directly correspond to the dynamical aspect of the genuine Rabi
problem~\cite{knight1980rab}. Rabi rotations instead reflect the
final exciton state and
thus lack direct visibility of the temporal interplay between
exciton and phonons.

We here propose and realize an original 
approach to monitor single exciton's Rabi oscillations in the time
domain. Namely, we use the delay dependence of four-wave mixing
(FWM) signals to probe the dynamics of both central quantum mechanical
quantities of a single QD exciton,
i.e., its occupation and its microscopic coherence,
which represent the entire quantum state or its Bloch vector.
Note that combined with the time-dependence of the emitted signal, this delay dependence can be used to generate
two-dimensional spectra. These provide a rich playground to study
spectral features of coupled and uncoupled quantum
systems~\cite{borca2005opt,cundiff2008coh,kasprzak2011coh,moody2013inf,cundiff2013opt,mermillod2016dyn,lomsadze2017fre}. Here, however, our focus is on the temporal evolution of the exciton quantum state. 
Our dual insight into Rabi oscillation
dynamics of a single QD exciton represents a novel
aspect and is enabled by the FWM methodology. In a combined
experimental and theoretical study we investigate the dynamics of both quantities in FWM signals. When increasing the intensities of the exciting laser pulses, oscillations on the ps time scale build up during the optical driving. The measured oscillations directly reflect Rabi oscillations
of the Bloch vector of an individual QD exciton. In exciton
ensembles the isolation of a single Bloch vector has so far not been
realized, because one always deals with a bunch of Bloch vectors
undergoing different dynamics.
Carrying out this study in the ultimate limit of a
single QD exciton allows for an investigation of the fundamental
interplay between the optically driven two-level system and its
coupling to the host lattice via the exciton-phonon interaction.

\section{EXPERIMENT AND THEORY}
\begin{figure}[h!]
\centering
\includegraphics[width=0.65\columnwidth]{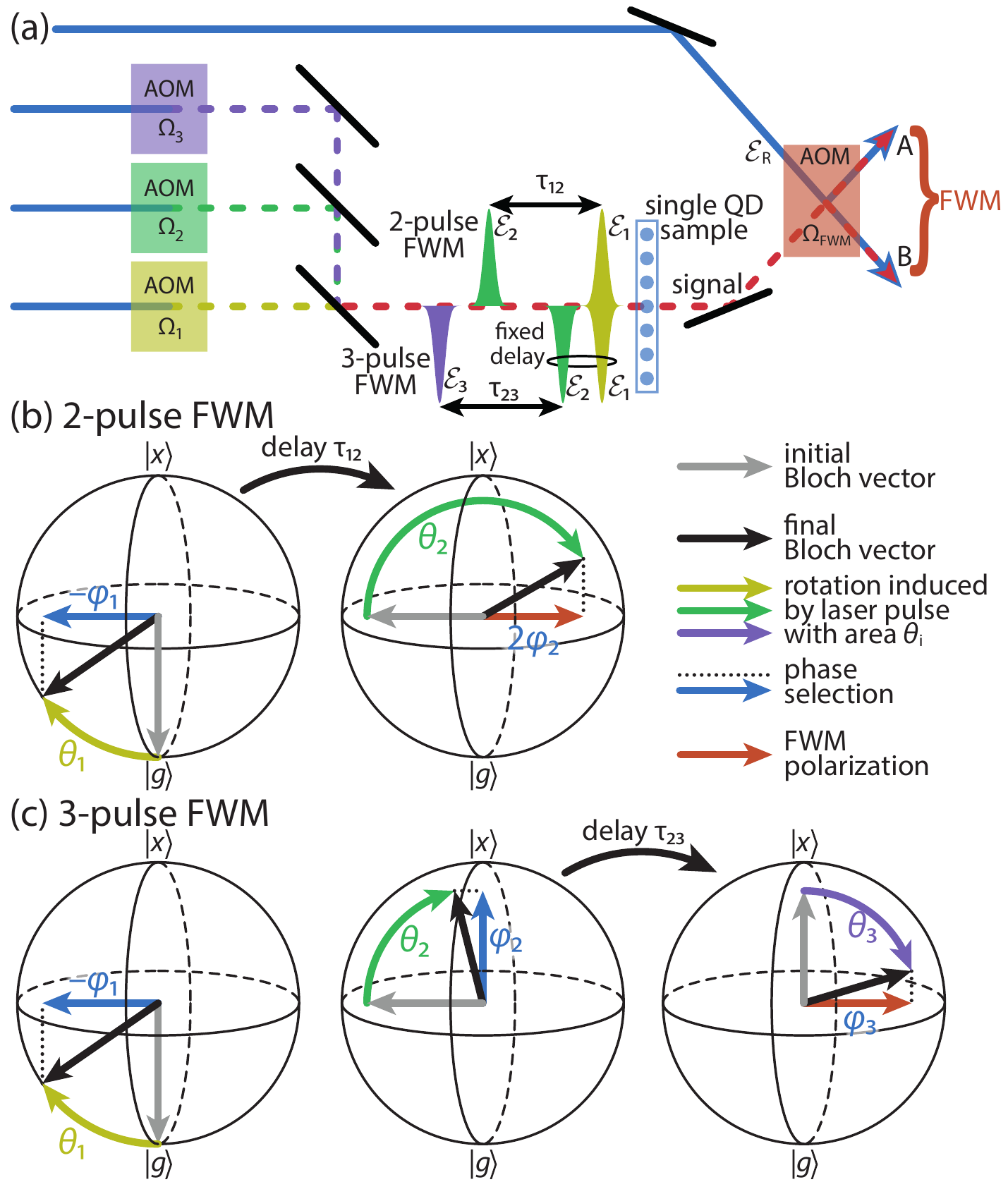}
\caption{Schematic picture of the experiment and the theory.
(a) Experimental setup. Acousto-optical modulators (AOMs) label the
excitation pulses with radio frequencies $\Omega_i$. The signal beam
after the sample is mixed in a AOM with the reference beam to generate
 the stationary spectral interference at the CCD camera, corresponding to the FWM signal. Note that the actual experiment is working in reflection
from the sample. To keep the picture as simple as possible the figure
shows transmission geometry. (b, c) Schematic picture of the theoretical
simulation of the exciton state on the Bloch sphere. After each pulse with
pulse area $\theta_i$ the state is filtered with respect to phase factor $\varphi_i$
corresponding to the final FWM phase $\varphi_{\rm FWM}$. (b) For
two-pulse FWM where $\tau_{12}$ is scanned and (c) for three-pulse FWM
where $\tau_{23}$ changes.} \label{fig:Fig0}
\end{figure}

To measure FWM on individual QDs, we employ a three-beam heterodyne
spectral interferometry setup optimized for the near-infrared
spectral range~\cite{mermillod2016dyn} as
schematically depicted in Fig.~\ref{fig:Fig0}. The FWM is driven
using a Ti:Sapphire femtosecond laser. The QD exciton is excited by Gaussian laser pulses, formed by a passive  pulse shaper based on a diffraction grating. The pulse train is:
i)\,split into three beams ${\cal E}_i$ ($i={1,2,3}$) and a
reference beam $\Er$, ii)\, the beams ${\cal E}_i$ are individually
phase-modulated using acousto-optic
modulators (AOMs)
operating at distinct radio-frequencies $\Omega_i$ around 80\,MHz,
iii)\,the pulses $\Eo$ are delayed with respect to one another by
$\tau_{12}$ and $\tau_{23}$, iv)\,their intensities $P_i\sim|{\cal
E}_i|^2$, which are varied to scan the pulse areas $\theta_i$, are
controlled via fixed radio-frequency drivers. After steps
i)--iv) $\Eo$ propagate 
co-linearly and are focussed on the
sample surface. The pulse shaper is also used to geometrically compensate the first-order chirp, attaining a focal spot close to the Fourier limit. FWM is retrieved in reflectance (note that,
for the sake of readability, Fig.~\ref{fig:Fig0}(a) shows a
transmission configuration) by locking $\Er$ at the specific
heterodyne frequency ($\Omega_{\rm FWM}=2\Omega_2-\Omega_1$ or
$\Omega_3+\Omega_2-\Omega_1$) and interfering it with the signal
beam from the sample (red in Fig.~\ref{fig:Fig0}(a)),
which is phase-modulated with the same heterodyne frequency as
$\Er$. The FWM signal is retrieved by subtracting the intensities in channels A and B from each other, attaining the shot-noise limited detection. $\Er$ is also used to generate background free spectral
interference on a CCD camera installed at the output of an imaging
spectrometer.

In these experiments, the $\Eo$ pulse durations, hereafter denoted
as $\tau$, arriving at the QD position are in a few hundred fs up to
ten ps range. Therefore, the theoretical treatment of the optical
excitation has to go beyond the computationally handy ultrashort-pulse
limit, where analytic expressions for the FWM signals can be
derived~\cite{kasprzak2013vec,mermillod2016dyn}. To simulate the FWM
signals of the optically driven QD we model the QD exciton as a two-level system coupled to LA phonons via the deformation potential mechanism~\cite{krummheuer2002the}. The equations of motion for the occupation and the polarization are then obtained with the well established
correlation-expansion approach~\cite{rossi2002the}. The full set of
equations of motion are given in Ref.~\cite{krugel2006bac}.
Practically, to compute the FWM signal we drive the QD exciton with
Gaussian laser pulses
\begin{equation*}
\mathcal{E}_i(t) = \frac{\hbar}{2\sqrt{2\pi}}\frac{\theta_i}{\mu\tau}
\exp\left[ -\frac12\left( \frac{t-t_i}{\tau} \right)^2-{\rm i}\omega_{\rm L}t + {\rm i} \varphi_i \right]\ ,
\end{equation*}
in general with $i=1,\,2,\,3$, the pulse areas $\theta_i$, the dipole moment
$\mu$ and the pulse duration $\tau$ at times $t_i$. The central
energy is $\hbar\omega_{\rm L}$, which is chosen to be in resonance
with the exciton transition. We label each
pulse with a phase $\varphi_i$ with $i=1,2$
for two-pulse and with $i=1,2,3$ for three-pulse FWM. These phases
$\varphi_i$ correspond to the radio-frequency shifts $\Omega_i$ in
our experimental heterodyne detection scheme. The FWM signal
$S_{\rm FWM}$ is then isolated by filtering the resulting microscopic
polarization $p=\left<\left|g\right>\left<x\right|\right>$ with respect to the
FWM phases $\varphi_{\rm FWM}=2\varphi_2-\varphi_1$ or
$\varphi_{\rm FWM}=\varphi_3+\varphi_2-\varphi_1$ resulting in the FWM polarization $p_{\rm FWM}\sim S_{\rm FWM}$. Here, $\left|g\right>$
denotes  the ground state and $\left|x\right>$  the exciton state. 
Note that this filtering of the signal with respect to the phases directly mimics the heterodyne detection scheme. In particular, although both in experiment and in theory the lowest order contribution to the FWM signal is of the order $\chi^{(3)}$, the signals are not limited to this order. Instead they include all nonlinear odd orders of the susceptibility $\chi$. This allows us to directly resolve Rabi oscillations both in experiment and in theory. By varying
the delay $\tau_{12}=t_2-t_1$ in two-pulse FWM and $\tau_{23}=t_3-t_2$
in three-pulse FWM we probe the coherence and population dynamics,
respectively. This can be seen from the Bloch sphere (spanned by the real and imaginary part of $p$ and the exciton occupation $f=\left< \left|x\right>\left< x\right|\right>$) schematics in Figs.~\ref{fig:Fig0}(b, c).
Due to the excitation with phase-modulated laser pulse train, for each repetition, the Bloch vector switches from its initial (grey) to its final state (black). The curved arrows represent the pulse areas of each driving pulse. The successive selection of the FWM phase after each pulse results in the final FWM polarization (orange arrow). In the experiment, optical heterodyning allows to probe a FWM representation of the Bloch vector dynamics in its rotating frame. Changing the delay
$\tau_{12}$ ($\tau_{23}$) probes the dynamics of the polarization (population)
in the case of two-pulse (three-pulse) FWM, as can be seen from the selected Bloch vector (dotted black line and blue arrow) directly before $\tau_{12}$ ($\tau_{23}$). The simulations reaching delay
times over 100~ps are computationally demanding, which makes finding
appropriate system parameters like pulse areas, pulse duration and QD
geometry quite challenging. Note, that we add a phenomenological long-time
dephasing rate $\beta$ for the polarization and a spontaneous decay
rate $\gamma$ for the exciton occupation to the model when
simulating dynamics in the 100~ps range.

\section{RESULTS}
\label{sec:results} We have recently shown that the pulse areas
$\theta_i$ of the driving laser fields in FWM experiments have a
strong influence on the system's dynamics between the pulses. For
instance, the visibility of quantum beats, arising due to the
splittings in the excitonic level structure of a QD system, depends
on the pulse area~\cite{mermillod2016dyn}. Furthermore, in a
two-pulse FWM experiment the strength of the phonon-induced
dephasing (PID) becomes more pronounced when increasing $\theta_1$
towards $\pi$~\cite{wigger2017exp}.

The pulse area dependence of the exciton quantum state is rather
obvious, when considering the dynamics during the interaction with
the laser field, i.e., Rabi oscillations take place. When keeping
$\tau$ fixed and increasing $P_i$ and therefore $\theta_i$, the
exciton is excited and de-excited during the pulse more often. In
the picture of the Bloch vector, it performs more rotations, which
means that the rotation speed, i.e., the Rabi frequency $\Omega_{\rm
R}$, increases. Additionally, the coupling between exciton dynamics
and phonons depends on the instantaneous Rabi frequency and the
phonon spectral density $J(\omega_{\rm ph})$. For the deformation
potential coupling between the QD exciton and LA phonons
$J(\omega_{\rm ph})$ scales like $\omega_{\rm ph}^3$ for small
phonon frequencies $\omega_{\rm ph}$~\cite{weiss2012qua}. For
typical self-assembled InGaAs/GaAs QDs with sizes in the range of a
few nm, the spectral density forms a broad maximum at $\omega_{\rm
ph,0}$ in the range of a few ps$^{-1}$, i.e., a few
meV~\cite{wigger2014ene}. The upper cut-off frequency is roughly
given by $\omega_{\rm ph,max}=2c/a$, where $c$ is the sound velocity
and $a$ the localization length of the exciton. Therefore the
strongest interaction between exciton and LA phonons lies in the
range of $\Omega_{\rm R}\approx \omega_{\rm ph,0}$.

\begin{figure}[h!]
\centering
\includegraphics[width=0.65\columnwidth]{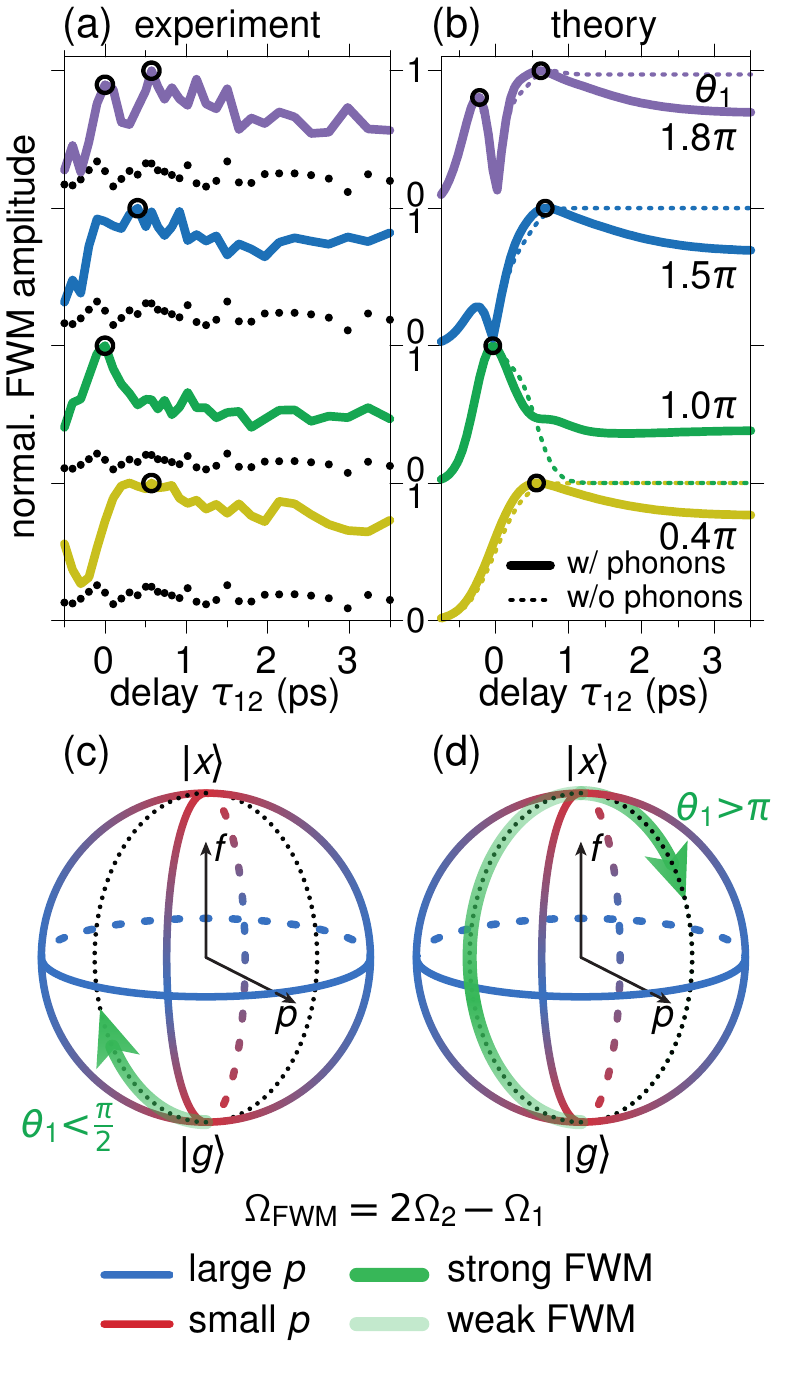}
\caption{Normalized two-pulse FWM generated with sub-ps pulses on a QD
exciton embedded in a low-Q cavity. (a, b) FWM amplitude as a function
of the delay $\tau_{12}$ for increasing pulse area of the first
pulse from bottom to top. The FWM delay dependence probes the optically driven evolution of the exciton polarization, performing Rabi oscillations for excitations with high pulse areas. (a) Experiment, the $P_1$ impinging the
sample surface are $(0.14,\,1.1,\,2.3,\,2.7)$~\textmu W,
$P_2=1$~\textmu W, temperature $T=23$~K. The dots indicate
the noise level. (b) Theory, solid lines with phonon coupling dotted
lines without phonons. $\theta_1$ as given in the plot. The circles mark maxima of the FWM signal.
(c, d) Schematic pictures of the FWM signal on the Bloch sphere.
(c) For small pulse areas $\theta_1$, (d) for pulse areas $\theta_1$
exceeding $\pi$.} \label{fig:Fig1}
\end{figure}

In the limit of ultrafast laser pulse excitation, in the range of
$\tau\approx 100$~fs, we have shown that simulations in the
delta-pulse limit yield a satisfactory agreement with
experiments\cite{mermillod2016dyn,jakubczyk2016imp}. In this limit,
the properties of the excited phonons only depend on the final
occupation of the exciton state~\cite{vagov2002ele}. Therefore, to sense the PID effects related with the
variation of the pulse area, one needs to work with longer pulses.
To this aim, we spectrally shape the initial laser beam, setting
durations of $\tau\simeq300$~fs. To enhance the influence of LA
phonons, we set the temperature to $T=23$~K, increasing the phonon
occupation of modes in the range of 7~meV by a factor of $7\times10^6$
with respect to 4.2~K, while keeping sufficiently long dephasing
of the zero-phonon line~\cite{borri2005exc,jakubczyk2016imp}. $T=23$~K
is also considered in the simulations. We
first present the results obtained on a neutral exciton in an InAs
QD embedded in a planar cavity, exhibiting a low quality factor
$Q_{\rm{planar}}\simeq1.7\times10^2$, as recently employed in
Refs.~\cite{Fras2016mul, mermillod2016dyn,wigger2017exp}. The layer of annealed and capped InAs QDs (density $2\times10^9$~cm$^{-2}$) is placed in the center of a GaAs spacer. The spectrum of the driving laser pulses is tuned to cover the ground state to exciton transition as shown in Fig. S1(a) in the Supplementary Material. Also the exciton to biexciton transition is slightly covered by the tail of the pulse spectrum, generating small signals for negative delays via the two-photon coherence (see Fig. S1(a)). A strong influence from the biexciton state would also lead to a beating for positive delays due to the biexciton binding energy~\cite{mermillod2016dyn}. Because the pulse spectrum has only little overlap with the exciton to biexciton transition, we do not resolve this beating for positive delays. Therefore, we conclude that the influence of the biexciton for the signals at positive delay times is negligible and neglect the biexciton state in our study, restricting it to a two-level system. In
Figs.~\ref{fig:Fig1}(a) and \ref{fig:Fig1}(b), we plot the two-pulse FWM amplitude
as a function of $\tau_{12}$. The FWM phase is given by
$\varphi_{\rm FWM}=2\varphi_2-\varphi_1$, meaning that the first
pulse creates a coherence and the second pulse converts this
coherence into the FWM signal (see Fig.~\ref{fig:Fig0}(b)). We show results for four different pulse areas
$\theta_1$ at a fixed second pulse area of $\theta_2=\pi$. Panel \ref{fig:Fig1}(a) presents
the measurement and panel \ref{fig:Fig1}(b) the respective simulation results for small to large
pulse areas from bottom to top. Note, that all curves are normalized to unity. The pulse areas in the calculations
are listed in the plot next to the respective graph. For the
geometry of the QD we choose slightly different parameters than in
Ref.~\cite{wigger2017exp}. Here, the electron and hole
localization lengths are $a_{\rm e}=7$~nm and $a_{\rm h}=1.5$~nm,
respectively. These parameters give the best agreement between the
measured data in Fig.~\ref{fig:Fig1}(a) and the simulations in \ref{fig:Fig1}(b).
Note, that the dimensions of the exciton $a_{\rm e}$ and
$a_{\rm h}$ are a spherical representations of the exciton, which leads to
the same physical results as a lens-shaped model~\cite{luker2017pho}.
Therefore, these sizes must not be seen as the real size of the
exciton wave function.

We start the discussion with the smallest measured pulse area at the
bottom in Fig.~\ref{fig:Fig1}(a) (yellow). The FWM amplitude builds up around $\tau_{12}=0$
and reaches a maximum at around 0.5~ps (marked by the circle). After
that, the signal slightly decays within the next 3~ps. This behavior
is well reproduced by the simulation in Fig.~\ref{fig:Fig1}(b) including the coupling to
LA phonons (solid) for $\theta_1=0.4\pi$. The drop of the signal
within the first few ps results from the PID effect. Together,
the phonons and the exciton in the QD form a new equilibrium
state, the acoustic polaron. This polaron is accompanied by a
static lattice displacement in the vicinity of the QD~\cite{hohenester2007qua}.
When the exciton, and therefore the polaron, is created faster than the
typical timescale of the involved phonons, a phonon wave packet is
emitted.  This results in an irreversible loss of exciton coherence~\cite{vagov2003imp,vagov2004non,jakubczyk2016imp}. To
emphasize that the signal drop results from the phonon coupling, we
also provide simulations neglecting the exciton-phonon interaction,
resulting in the dashed yellow curve in Fig.~\ref{fig:Fig1}(b),
visibly lacking the initial decay due to the PID effect.

Going over to the next larger pulse area in Fig.~\ref{fig:Fig1}(a) (green), we clearly
see that the drop of the measured FWM signal is significantly
increased and reaches an almost stationary value within less than
1~ps. Additionally, the circle shows that the maximum of the signal
shifts to shorter delays. These findings are consistent with the
simulation for $\theta_1=1.0\pi$ in panel (b) (solid green). However, the
calculated FWM signal without the exciton-phonon coupling (dashed
green) goes exactly to zero after $\tau_{12}=1$~ps. To explain this
we take a look at the schematic Bloch sphere picture in
Fig.~\ref{fig:Fig1}(c), where the green arrow visualizes the
movement of the Bloch vector during the first laser pulse
excitation. When the first pulse area is small, i.e.,
$\theta_1<\pi/2$,  the microscopic polarization $p$ is maximal at
the end of the pulse. Therefore, also the two-pulse FWM signal is
maximal for small positive delays, i.e., after the first pulse. This
is the case for the dashed yellow curve. However, for pulse areas
$\pi/2<\theta_1<\pi$ the Bloch vector crosses the equator of the Bloch sphere, i.e., the
state of maximum coherence, during the pulse. For the special case
of a $\pi$-pulse, the final polarization is zero, leading also to a
vanishing FWM signal. Already here we see
that the dynamics of the FWM signal resembles more involved movements of the
Bloch vector during the action of the first laser pulse. The full model, including the coupling to LA
phonons, is shown as solid green line in Fig.~\ref{fig:Fig1}(b). We
see that the stationary value of the curve strongly deviates from
the dashed line. This mismatch is again a consequence of the PID
effect. Going from a $0.4\pi$ pulse to a $\pi$-pulse excitation the
dephasing influence of the phonon coupling should significantly
increase because the final exciton occupation $f$ is approximately twice
as large~\cite{wigger2013flu,wigger2014ene}. Comparing the solid lines with
the dashed lines of the same color (yellow and green), we see that
this is clearly the case. A stronger excitation leads to a larger
phonon wave packet and therefore to stronger dephasing.

When increasing the pulse area further, in the second curve from
the top in Fig.~\ref{fig:Fig1}(a) (blue), the signal drop decreases
again and the maximum of the signal also goes back to longer
$\tau_{12}\approx0.5$~ps as marked by the circle. We find the best
agreement with the experiment for the simulation with
$\theta_1=1.5\pi$ in panel (b) (solid blue). The dynamics of the
simulated signals, both with and without the exciton-phonon
interaction, look very much like the $\theta_1=0.4\pi$ case
(yellow). The only significant difference of the blue curve is an
additional minimum at $\tau_{12}=0$~ps. This is again a result of
the Bloch vector dynamics. We are now dealing with the
$\theta_1>\pi$ case, schematically shown in Fig.~\ref{fig:Fig1}(d).
The Bloch vector crosses the north pole of the sphere during the
pulse, which leads to null polarization and in consequence to a
vanishing FWM signal. After 1~ps the solid blue line in panel (b) is
again governed by the PID drop of the signal. The additional minimum
of the FWM signal around $\tau_{12}=0$~ps is obviously too
unpronounced to be clearly resolved in the experiment in panel (a).
When comparing the simulation with and without phonon coupling we
find that the drop of the signal, i.e., the PID effect, gets weaker
with respect to the pulse area discussed above. This is in line with
the previous explanation of the strength of the PID effect. For this
pulse area the final exciton occupation is smaller than for the
$\pi$-pulse, which leads to weaker dephasing.

When we move to the largest considered pulse area at the top
(violet) a clear minimum shows up in the measured signal in
Fig.~\ref{fig:Fig1}(a). This is in excellent agreement with the
simulation for $\theta_1=1.8\pi$ in panel (b) (solid violet). Here,
the signal minimum is significantly more pronounced than in the case
considered before. This is an instructive demonstration of the
optically driven Rabi oscillations of the exciton state. The PID
drop after $\tau_{12}=0.5$~ps is of a comparable strength as for the
$1.5\pi$ case in blue.

From the pulse area series in Figs.~\ref{fig:Fig1}(a) and
\ref{fig:Fig1}(b) we find that the effect of the PID and therefore
the influence of the exciton-phonon interaction changes measurably
with the pulse area $\theta_1$ of the first driving laser pulse. For
large positive delays, i.e. $\tau_{12}>\tau$, pulse one arrives
first, whilst for large negative delays pulse two arrives first.
Therefore, around $\tau_{12}=0$ both pulses overlap. The additional
dynamics evolving around $\tau_{12}=0$ happen during the presence of
the first pulse. Because the $\tau_{12}$-dependence of the two-pulse
FWM signal represents the dynamics of the coherence, i.e., the
microscopic polarization of the exciton $p$, the resolved
oscillations for the largest considered pulse area stem from
optically driven Rabi oscillations in the time domain.

It was shown that the exciton-phonon interaction leads to more
involved movements of the Bloch vector during optical driving. The phonon lead to i) dephasing, i.e., shrinking of the Bloch vector length
and ii) a mixture of real and imaginary part of the
polarization~\cite{barth2016fas}, i.e., a movement out of the Rabi oscillation
plane of the Bloch vector. For simplicity, the illustrating pictures
in Figs.~\ref{fig:Fig1}(c) and \ref{fig:Fig1}(d) do not take these complexities into
account. Instead, they capture well the origin of the Rabi
oscillations observable in the FWM delay dynamics. Following this
proof of principle demonstration, we now generalize this approach
using a more suited photonic nanostructure.

\begin{figure}[h!]
\centering
\includegraphics[width=0.7\columnwidth]{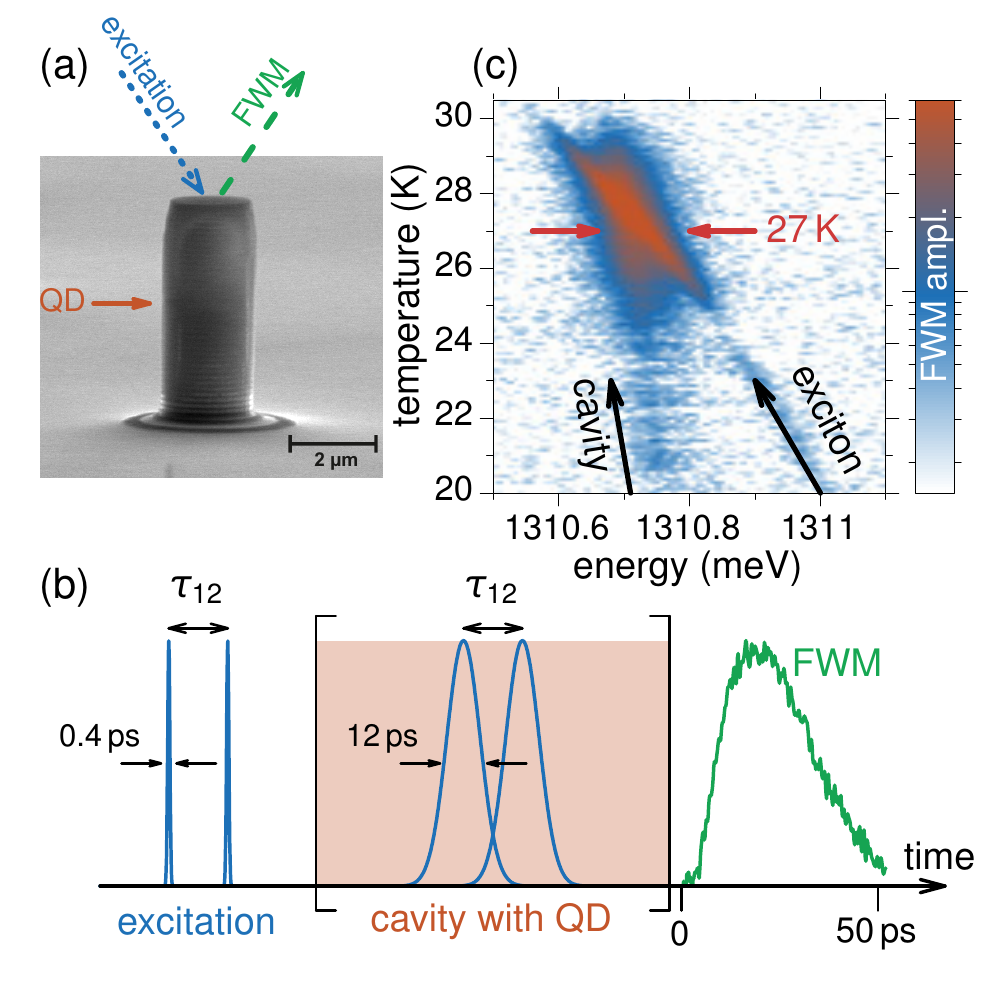}
\caption{Micro-pillar cavity system. (a) Scanning electron
microscopy image of an exemplary micro-pillar cavity system with a diameter of 1.8~\textmu m and a height around 10~\textmu m. (b) FWM spectra for varying
temperatures, demonstrating operation in the weak-coupling regime:
exciton and cavity resonances shift in energy and cross at $T\approx
27$~K. (c) Schematic picture of the driving laser pulses and the
measured FWM dynamics. The effective pulse duration $\tau$ is
increased by a factor of 30 inside the cavity. Green trace is the
measured time-resolved FWM field (vertical logarithmic scale)
illustrating its build up owing to the high Q-factor.}
\label{fig:Fig2}
\end{figure}

Shifting the focus from the PID phenomenon to the Rabi oscillations,
we aim to drive the QD with laser pulses that are as long as
possible, to stretch the dynamics in time. In parallel, we want to
reach preferably large pulse areas, to generate as many Rabi periods
as possible. Our FWM methodology relies on the detection of
spectrally-resolved interference between the QD emission and the
reference $\Er$, requiring a bandwidth well beyond the
spectrometer resolution, in practice over 0.1~meV. As such, the
experiment is limited to pulse durations $\tau$ in the few ps range.
To overcome this issue, we switch to a micro-pillar
cavity~\cite{reithmaier2004str,Kasprzak2010upo} with a diameter of 1.8 \textmu m and a height around 10 \textmu m, as depicted in Fig. 3(a), which yields a two orders of magnitude higher
quality factor of $Q_{\rm{pillar}}\simeq2.4\times10^4$. This structure, containing a layer of
InAs QDs at the antinode of the cavity mode, allows us to access the
photonic structure with sub-ps pulses (namely, $\tau\simeq0.4$~ps). See Supplementary Material for spectra of  spectral interference and laser pulses.
We tightly focus them onto the top facet of the pillar and construct
spectral interferences with \Er, placed 12~\textmu m apart at the
auxiliary pillar~\cite{Kasprzak2010upo}. Crucially, due to the high
quality factor the optical field reaches the QD in the cavity in a
retarded way. This stretches the pulses to durations of over 10~ps,
as schematically shown in Fig.~\ref{fig:Fig2}(b).
Figure~\ref{fig:Fig2}(c) shows a temperature scan of the FWM
spectrum of the QD-cavity system (a corresponding scan of the
photoluminescence intensity is shown in the Supplementary Material).
The resonances of exciton and cavity are marked in the picture.
Around $T=27$~K both resonances cross, which shows that the coupled
QD-cavity system operates in the weak coupling
regime~\cite{reithmaier2004str,yoshie2004vac,peter2005exc}. The
possibility of adjusting the detuning very precisely makes this
system prototypical to demonstrate the Purcell
effect~\cite{purcell1946spo,gerard2001ina}. We do so by applying a
three-pulse FWM measurement for different detunings between the
cavity mode and the exciton transition. The results are shown in the
Supplementary Material.

The enhancement of the intra-cavity field in the pillar structure
allows for reaching larger $\theta_i$ with respect to the low
Q-factor planar cavity explored in Fig.\,\ref{fig:Fig1}. In the
following, we choose the temperature to $T=27$~K, setting the
exciton transition to be approximately in resonance with the cavity
mode as marked by the red arrows in
Fig.~\ref{fig:Fig2}(c). Working at these elevated temperatures
results in a significant increase of the LA phonon influence than at
5~K~\cite{jakubczyk2016imp}. In all simulations we choose the
temperature to 25~K.

\begin{figure}[h!]
\centering
\includegraphics[width=0.7\columnwidth]{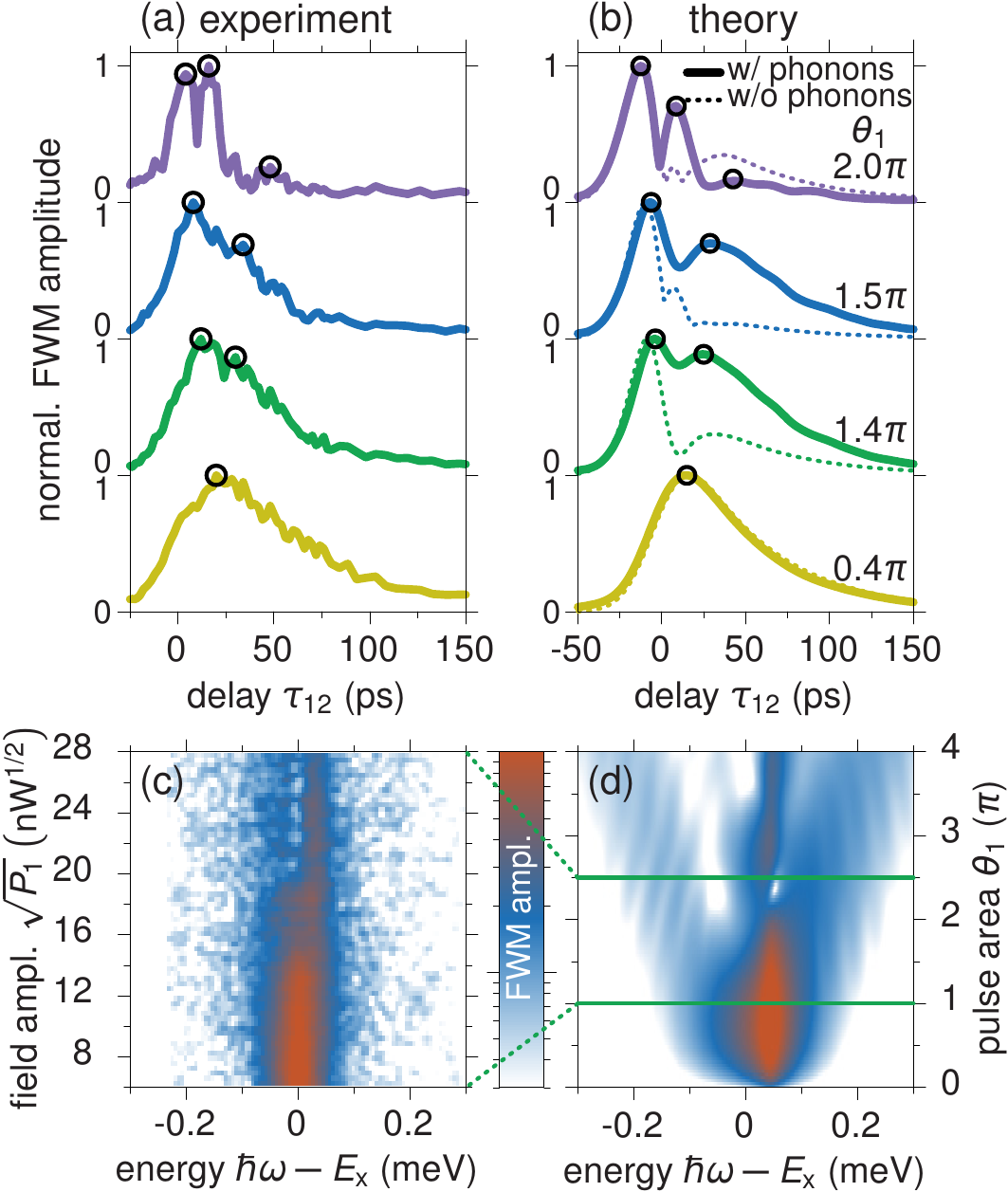}
\caption{Normalized two-pulse FWM with
$\tau\simeq12$~ps pulses in a micro-pillar cavity. (a, b)~FWM
amplitude as a function of the delay $\tau_{12}$ for increasing
pulse area of the first pulse from bottom to top. (a)~Experiment,
$P_1=(0.02,\,0.24,\,0.35,\,0.55)$~\textmu W, $P_2=0.08$~\textmu W.
(b)~Theory with $\theta_1$ as given in the plot and $\theta_2 =
\pi$. Solid/dotted lines with/without coupling to phonons. (c,
d)~FWM spectral amplitudes as a function of excitation power for
$\tau_{12}=0$ illustrating emergence of the Autler-Townes splitting
with increasing $\theta_1$. Experiment in (c) against $\sqrt{P_1}$
and theory in (d) against $\theta_1$.} \label{fig:Fig3}
\end{figure}

The results for the two-pulse FWM study are shown in
Fig.~\ref{fig:Fig3} with the measurements in panel (a) and the
simulations in panel (b). All curves are normalized to their respective 
maximum. We consider four different pulse areas,
increasing from bottom to top. The second pulse area is fixed in the
simulations to $\theta_2=\pi$. The pulse duration $\tau$ and the
long time dephasing rate $\beta$ are fitted to the smallest pulse
area and we found the best agreement for $\tau=12$~ps and
$\beta=0.01$/ps. For the simulations we choose the same QD
geometry as for the calculations shown in Fig.~\ref{fig:Fig1}.

For the smallest considered pulse area the measured FWM signal in Fig.~\ref{fig:Fig3}(a)
just forms one maximum around $\tau_{12}=25$~ps and decays for
longer delays single exponentially with $\beta$. These dynamics are
well reproduced by the simulation in Fig.~\ref{fig:Fig3}(b). Additionally, there is
nearly no difference between the calculation with and without
exciton-phonon interaction (solid and dashed). This indicates that
the optically driven dynamics are to slow to induce the PID drop of
the signal. In other words, the polaron creation follows the exciton
occupation adiabatically and no phonon wave packet is
emitted~\cite{wigger2014ene}. We can instructively describe the optical
excitation process using the laser-exciton dressed state picture~\cite{tannor2007int}.
These states then have an energy splitting $\Delta_{\rm AT}$, the so
called Autler-Townes (AT) splitting~\cite{autler1955sta}, which is proportional
to the instantaneous laser field amplitude. While in the exciton
basis the LA phonons only couple to the exciton state, in the presence of a light field they lead to
transitions between the dressed states. For
small pulse areas, i.e. $\theta_1\ll\pi$, $\Delta_{\rm AT}$ is
correspondingly small, in the ~\textmu eV range, hence in the energy range
where the phonon spectral density is negligible~\cite{wigger2017exp}.
Therefore, the phonons do not lead to efficient transitions between the
dressed states, which allows them to evolve adiabatically.

The appearance of a splitting between the dressed states with increasing field strength
can be seen in Figs.~\ref{fig:Fig3}(c) and \ref{fig:Fig3}(d). There, we show the
FWM spectrum centered around the
exciton transition energy $E_{\rm x}$. In the measurement in panel (c) it is plotted
against $\sqrt{P_1}\sim \theta_1$, which is proportional to the
field amplitude. The corresponding simulation is presented in panel (d) and
shows a good agreement with the measurement.
The spectra are taken for $\tau_{12}=0$. While for small pulse areas
the spectrum is dominated by a single line at the exciton energy, it
splits according to the AT splitting of the dressed states 
reaching values of approximately 0.1~meV in panel (c). 
Note, that the AT splitting is time dependent for pulsed excitations. Therefore, we are seeing a time integrated version of the splitting between the dressed states in the FWM spectra. However,
$\Delta_{\rm AT}$ is the direct spectral domain translation of the
Rabi oscillations in the FWM signal. In agreement between the
dynamics depicted in Figs.~\ref{fig:Fig3}(a) and \ref{fig:Fig3}(b) and the
spectral behavior in Figs.~\ref{fig:Fig3}(c) and \ref{fig:Fig3}(d), we
find clear signatures of the Rabi oscillations from pulse areas
of $\theta_1\approx 1.5\pi$ onward. We additionally find, that the
line at positive detuning is stronger than the other one. This stems
from a slight mismatch between the energies of the driving laser
pulses and the exciton transition of $\hbar\omega_{\rm L}-E_{\rm x}=
-0.05$~meV, which was considered in the simulation in Fig.~\ref{fig:Fig3}(d). This detuning
agrees well with the separation between the exciton and cavity line
in Fig.~\ref{fig:Fig2}(b) at 27~K. Only for the case of exact resonance
between the driving field and the transition energy the dressed states
are equally occupied~\cite{tannor2007int}. As we find here, already
detunings in the few \textmu eV range result in a significant mismatch
of the two lines in the FWM signal. To check that the AT splitting is driven
by the laser pulses we performed the same measurement and simulation
as in Figs.~\ref{fig:Fig3}(c) and \ref{fig:Fig3}(d) but with a large delay of
$\tau_{12}=30$~ps to reduce the overlap of the pulses. Here, no
AT splitting was found (see Supplementary Material).

Comparing the second pulse area in Fig.~\ref{fig:Fig3}(a) (green) to the
yellow curve, the maximum of the signal splits into two local
maxima, with one moving to smaller and one moving to larger delays
$\tau_{12}$. The maxima are marked by the circles. In the simulation
in Fig.~\ref{fig:Fig3}(b) we find the best agreement for
$\theta_1=1.4\pi$ (solid green line) where we already find clear signatures of the Rabi oscillations
during the interaction with the first laser pulse. Here two distinct maxima of almost
equal height are formed. Compared to the simulation without the
exciton-phonon interaction (dashed green line) we find a significant
deviation for the second maximum already for this pulse area.
Without the coupling to phonons, the FWM signal drops very rapidly
and the second maximum is much smaller than the first one. There are
two main effects enhancing the discrepancy between the two
calculations: i) For these pulse areas we have shown that
$\Delta_{\rm AT}$ attains the 0.1~meV range. This allows for more
efficient phonon assisted transitions between the dressed states and
therefore to strong dephasing during the first laser pulse. ii)~The
exciton-phonon interaction leads to a renormalization of the pulse
areas~\cite{forstner2003pho,vagov2007non,ramsay2010pho,glassl2011inf}. This makes a direct comparison of the pulse areas
with and without phonon coupling difficult.

Stepping to the next larger pulse area in Fig.~\ref{fig:Fig3}(a),
i.e., from the green to the blue curve, does not alter the dynamics
of the FWM signal significantly. We basically  find a slightly
larger difference in the height of the two maxima in the signal.
This is reproduced by the modeled signal in panel~(b). Here the minimum
between the two maxima is more pronounced as in the experiment,
which we already found in a similar way for the shorter pulses in
Fig.~\ref{fig:Fig1}. The deviation between the simulation with and
without phonon coupling remains remarkable because of the reasons
pointed out earlier.

The most significant difference for the signal dynamics is found for
the largest considered pulse area in the experiment, which is shown
as violet curve in Fig.~\ref{fig:Fig3}(a). It forms a double peak
structure within the first 25~ps followed by a minor maximum around
$\tau_{12}=50$~ps. After that the signal is basically null. The very
same behavior is found in the simulation with $\theta_1 = 2.0\pi$ in
Fig.~\ref{fig:Fig3}(b): Two strong and narrow peaks around
$\tau_{12}=0$~ps are followed by a small maximum at
$\tau_{12}\approx 50$~ps. Together with the stunning agreement with
the simulated curve in panel (b) (solid violet), this impressively shows
that multiple Rabi oscillations are resolved in the coherence
dynamics of the two-pulse FWM experiment. Without considering the
coupling to phonons the model gives the dashed line in
Fig.~\ref{fig:Fig3}(b). Here also three maxima build up, but the
relative heights of the second and third maximum do not agree with
the measured curve in panel (a) at all. This shows that coupling to phonons
has a strong impact on the optically driven dynamics of the exciton quantum state.

\begin{figure}[h!]
\centering
\includegraphics[width=0.7\columnwidth]{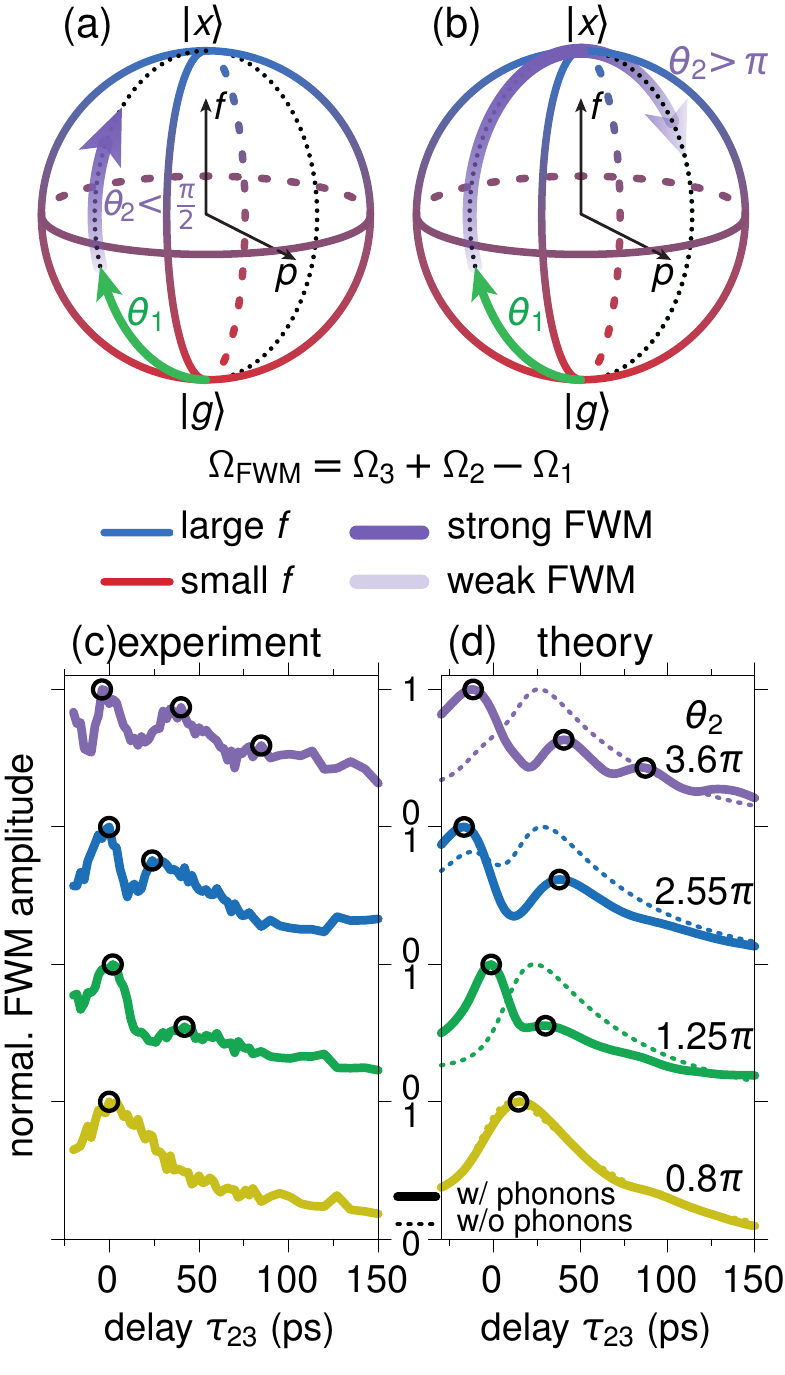}
\caption{Normalized three-pulse FWM with $\tau=12$~ps pulses
in a micro-pillar cavity. (a, b)~Schematic pictures of the FWM
signal on the Bloch sphere. (a)~For small pulse areas $\theta_2$,
(b)~for pulse areas $\theta_2$ exceeding $\pi$. Solid/dotted lines with/without coupling to phonons.
(c, d)~FWM amplitude
as function of the delay $\tau_{23}$ for increasing pulse area of
the second pulse from bottom to top. The FWM delay dependence probes the optically driven evolution of the exciton occupation, performing Rabi oscillations for excitations with high pulse areas. (c)~Experiment, $T=27$~K,
$\tau_{12}=10$~ps, $P_1=0.05$~\textmu W, $P_3=0.1$~\textmu W,
$P_2=(0.13,\,0.3,\,0.56,\,1)$~\textmu W. (d)~Theory with
$\theta_1=0.2\pi$, $\theta_3=0.4\pi$ and $\theta_2$ as given in the
plot.}
\label{fig:Fig4}
\end{figure}

The FWM technique allows to use different pulse sequences that lead
to different microscopic quantities determining the FWM signal. So
far we have studied the exciton polarization $p$ by employing a
two-pulse FWM experiment with $\varphi_{\rm FWM}=2\varphi_2-\varphi_1$.
We now turn to a three-pulse excitation with $\varphi_{\rm
FWM}=\varphi_3+\varphi_2-\varphi_1$ where - in a two level system -
the FWM signal carries information about the occupation of the
exciton state $f$. A schematic
picture of the Bloch sphere for this experiment is shown in
Figs.~\ref{fig:Fig4}(a) and \ref{fig:Fig4}(b) (see also Fig.~\ref{fig:Fig0}(c)). The first pulse drives a microscopic
coherence. The second pulse converts this coherence into an occupation
of the exciton, which is then turned into the FWM signal by the third
pulse. A larger occupation of the exciton state $f$ will accordingly lead
to a stronger FWM amplitude. The idea is to fix the first and third pulse
area to small values $\theta_1,\,\theta_3\lesssim\pi/2$ and change the
second area $\theta_2$. This will then result in different occupations
$f$ and therefore different signal strengths. For large pulse areas
$\theta_2$ the three-pulse FWM signal will resolve the optically driven
Rabi oscillations projected on the exciton occupation $f$.

Figures~\ref{fig:Fig4}(c) and \ref{fig:Fig4}(d) present the normalized experimental and
theoretical results, respectively, for different pulse areas
$\theta_2$ increasing from bottom to top. The FWM amplitude is
plotted as a function of the delay $\tau_{23}$ to investigate the
exciton population dynamics. The first delay is fixed to
$\tau_{12}=10$~ps to enable the build up of the intra-cavity field. We
see for the smallest pulse area at the bottom (yellow) that the
dynamics of the three-pulse FWM amplitude forms a single broad
maximum. For long delays the signal is dominated by the spontaneous
decay. The decay rate was chosen to $\gamma=0.004$/ps to give the
best fit to the measurement. We found the best agreement between
measurement and simulation for $\theta_2=0.8\pi$. The other pulse
areas are chosen to $\theta_1=0.2\pi$ and $\theta_3=0.4\pi$ in all
simulations.

Comparing experiment and theory we find
an excellent agreement for each considered $\theta_2$. With
increasing pulse strength, the first maximum shifts to shorter
delays $\tau_{23}$, whilst at longer $\tau_{23}$ additional maxima
emerge, as marked by the circles. These dynamics now happen during the
interaction with the second laser pulse. Therefore the signal is
dominated by the Rabi oscillations of the exciton occupation $f$.
For the two largest pulse areas (blue and violet) up to three distinct maxima can be
found in the signal. When comparing the simulations including the
exciton-phonon coupling (solid) with those omitting it (dashed), we
find clear qualitative differences. In the $\theta_2=1.2\pi$ case
(green) the maxima have a significant mismatch in their
delay $\tau_{23}$ of approximately 25~ps. Another striking example
is the largest pulse area at the top, where the full model including
the phonon coupling develops multiple modulations, while the
uncoupled calculation yields a single broad peak. One obvious reason
for this stronger discrepancy between the model with and without
phonons in Fig.~\ref{fig:Fig4}(b) compared to the results in
Fig.~\ref{fig:Fig3}(b) is the fact that the former one includes
three pulses, while the latter one only two. Therefore, the impact
of dephasing and pulse area renormalization come into play for one
more pulse, further enhancing their influence. Additionally, we
reach higher pulse areas, which results in general in a more
efficient exciton-phonon interaction.

\begin{figure}[h!]
\centering
\includegraphics[width=0.7\columnwidth]{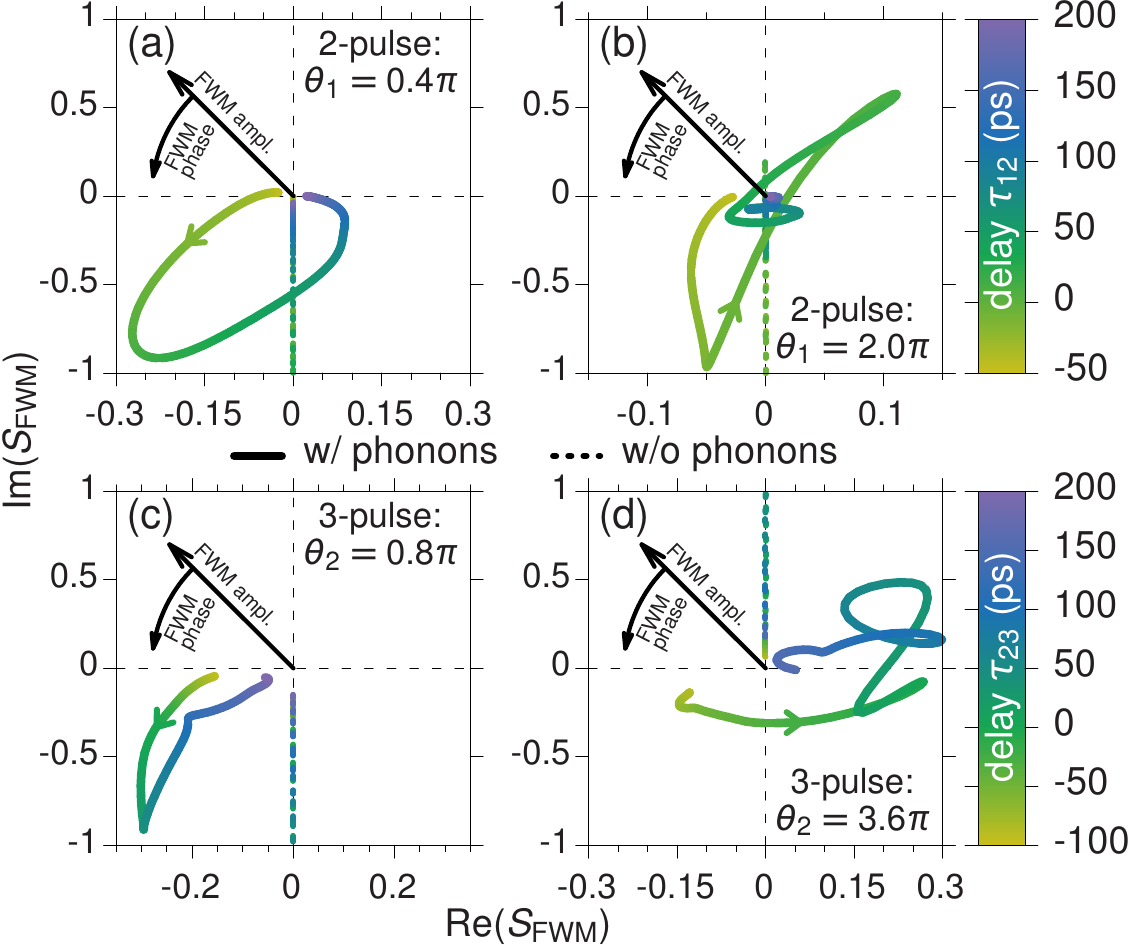}
\caption{Entire complex FWM signal. Real and imaginary part
of the FWM signal $S_{\rm FWM}$. The delays are color-coded,
the dotted lines show simulations without phonon coupling and
the solid lines with phonon coupling. (a, b) For 2-pulse FWM. The
corresponding dynamics of the FWM amplitude is given in
Figure~\ref{fig:Fig3}b. (c, d) For 3-pulse FWM. The
corresponding dynamics of the FWM amplitude is given in
Figure~\ref{fig:Fig4}d.}
\label{fig:Fig5}
\end{figure}

The FWM signal $S_{\rm FWM}$ is a complex quantity with real
and imaginary part or amplitude and phase. This feature has e.g.
been used to distinguish between different coupling situations in
few level systems~\cite{kasprzak2011coh}. We here use it to further illustrate the
dynamics of the Rabi oscillations and especially the influence of the
exciton-phonon coupling. In Fig.~\ref{fig:Fig5} we plot real and
imaginary parts of the simulated FWM signals and color-code the delay
to visualize the dynamics of the signals. The arrow heads indicate the direction of the time evolution. In principle, panels (a,\,b) and (c,\,d) are more
sophisticated representations of the data shown in Figs.~\ref{fig:Fig3}(b)
and \ref{fig:Fig4}(d), respectively. In panels (a) and (b) we show the
smallest and largest pulse area for the two-pulse FWM case and in panels
(c) and (d) the smallest and largest pulse areas for the three-pulse FWM
case. The remaining four considered pulse areas are given in the
Supplementary Material. The simulations without exciton-phonon
coupling are shown as dotted lines. These curves are restricted to the
vertical axis, i.e., the FWM signal is purely imaginary. When the full
model is considered (solid lines) real and imaginary parts of the FWM signals
get mixed and the curves in Fig.~\ref{fig:Fig5} show very involved
dynamics. While for the small pulse area examples in panels (a) and (c) the
signals are governed by a single loop, in the high area cases in panels (b) and
(d) the signals perform multiple loops. These loops represent the
Rabi oscillations, which we also found in the dynamics
of the FWM amplitudes in Figs.~\ref{fig:Fig3} and \ref{fig:Fig4}.
This way of analyzing the FWM signals
exploits the full potential of the method in an appealing fashion.

\section{CONCLUSIONS}
In summary, we have investigated the optically driven dynamics of
the full quantum state of a single QD exciton in the presence of
efficient coupling to acoustic phonons. By driving the system with
sub- and super-ps laser pulses and applying two- and three-pulse FWM
techniques, we could isolate the dynamics of the microscopic
polarization and the exciton occupation, respectively. By this we showed
that for large pulse areas involved dynamics of the Bloch vector in
the form of Rabi oscillations take place. Especially, the coupling
to LA phonons plays a decisive role for the optical control of the
QD exciton: On the one hand, for short laser pulses ($\tau<1$~ps)
phonons lead to the loss of coherence after the first laser pulse,
due to the emission of phonon wave packets. On the other hand, for
$\tau\simeq10$~ps the dephasing happens already during the first
pulse and it therefore strongly depends on the applied pulse area. Only
the accurate theoretical description of the exciton-phonon
interaction allowed for a quantitative analysis and comprehension of
the measured data. In the spectral domain, the observed Rabi
oscillations represent the Autler-Townes splitting, which we clearly
observed in the FWM spectra.

We thus find that FWM is a versatile technique to study optically
driven dynamics in many different aspects not only in single QDs but
also in any other potential isolated quantum systems like spins of
QD trions~\cite{Mi2018aco,Press2008com,DeGreve2011ult} or single defect centers in
insulators~\cite{Gao2015coh,Zhou2017coh}. However, the timescales of spins are by a
factor of 10~--~100 slower than our optically active exciton
transition, which renders the investigation in the optical range
far more challenging.

We have demonstrated that FWM spectroscopy of single emitters
can be used to study the optically driven dynamics of the quantum state
and at the same time works in a regime where the coupling to phonons
is rather strong. This makes this technique also promising to
investigate optical transitions in localized excitons in atomically thin
structures like transition metal dichalcogenides and hexagonal boron nitride or color centers in insulators. These
systems stand out due to their functionality at elevated temperatures
making them an up-and-coming platform for quantum applications.

\section{ACKNOWLEDGEMENTS}
We acknowledge the financial support by
the European Research Council (ERC) Starting Grant PICSEN (grant no.
306387). We further thank Sebastian L\"uker for support with the
implementation of the phonon coupled model.


\newpage
\renewcommand\thefigure{S\arabic{figure}}
\setcounter{figure}{0}

\section*{\large Supplemental Material}

\section*{1. Pulse spectra}
\begin{figure}[h!]
\centering
\includegraphics[width=0.6\columnwidth]{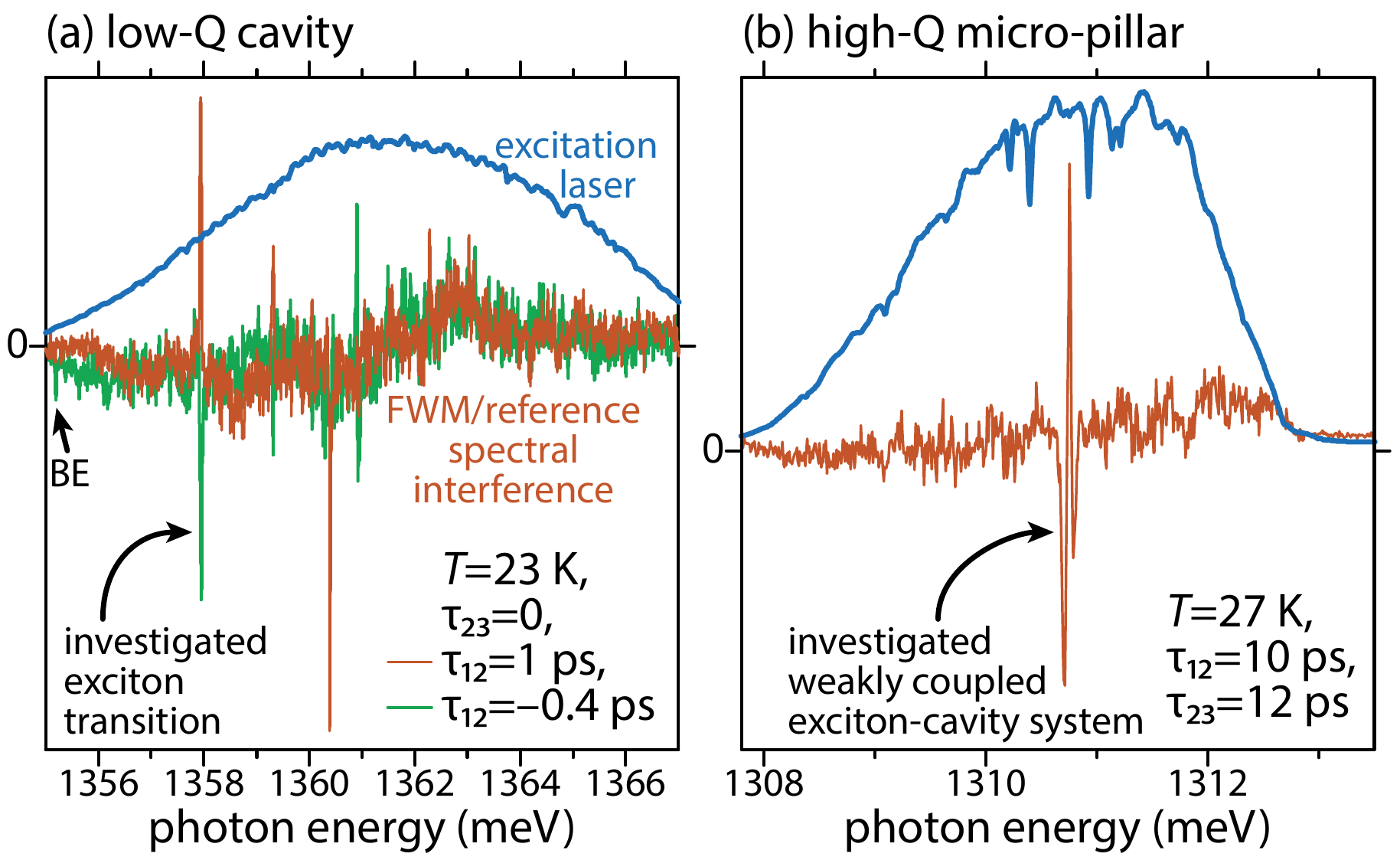}
\caption{Exemplary FWM spectral interferograms (orange and green traces) and spectra of the excitation laser (blue traces) measured on the low-Q planar micro-cavity (a) and high-Q micro-pillar (b). In (a) the exciton to biexciton transition is marked by (BE).}
\label{fig:pulses}
\end{figure}

Figure \ref{fig:pulses} shows four wave mixing (FWM) spectral interferograms (orange and green) of the low-Q cavity system studied in the first part of the paper in (a) and the weakly coupled high-Q micro-pillar cavity system in (b), which is studied in the second part. Spectra of the respective reference pulses are shown in blue. The spectral width of the laser in (a) is matched to the cavity mode. Extra dips in the laser spectrum in (b) are caused by the strong water absorption within that spectral range. Spectral and temporal shapes of the excitation attaining the quantum dot layer in (b) is defined, not by the external excitation laser,  but by the spectrally narrow cavity mode (around 50~\textmu eV), governing the intracavity field and strongly spectrally filtering the impinging laser.
~\\[-10mm]
\section*{2. Photoluminescence spectra}
\begin{figure}[h!]
\centering
\includegraphics[width=0.45\columnwidth]{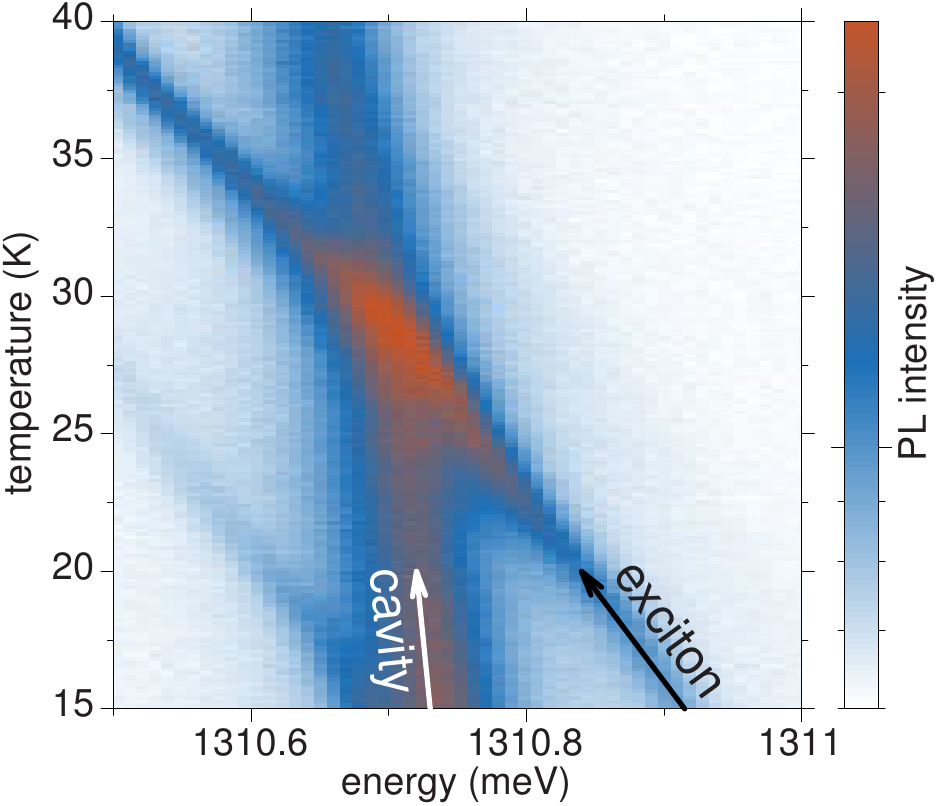}
\caption{Temperature scan of the photoluminescence spectra.}
\label{fig:PL_Temp}
\end{figure}

Figure \ref{fig:PL_Temp} shows photoluminescence (PL) spectra of the quantum dot micro-pillar cavity system considered in the main text for varying temperature. With this we confirm that the exciton and the cavity mode do not form an anticrossing around their resonance. This proves that we are operating the system in the weak coupling regime. A corresponding temperature scan of the FWM signal is given in Fig.~2(b) in the main text.

\section*{3. Autler-Townes splitting}
\begin{figure}[h]
\centering
\includegraphics[width=0.6\columnwidth]{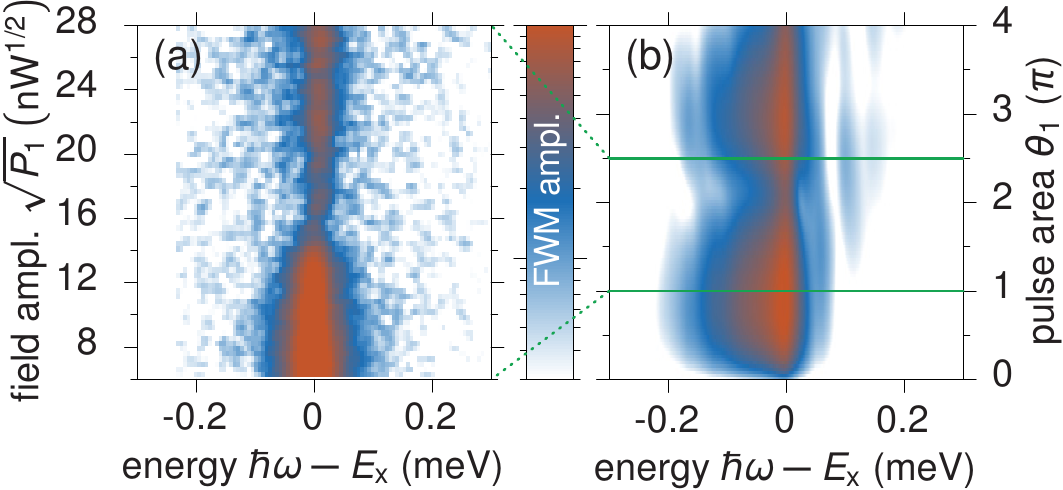}
\caption{Absence of the Autler-Townes splitting for large pulse delays. FWM spectra for different pulse areas at a delay of $\tau_{12}=30$~ps. (a) Measurement and (b) simulation.}
\label{fig:AT_SI}
\end{figure}

In the main text in Figs.~3(c) and 3(d) we show two-pulse FWM spectra for different pulse areas $\theta_1$ at the pulse delay $\tau_{12}=0$. There we show that for large pulse areas the exciton transition splits into the two dressed states separated by the Autler-Townes (AT) splitting. In Figs.~\ref{fig:AT_SI}(a) and \ref{fig:AT_SI}(b) corresponding spectra are shown for a delay of $\tau_{12}=30$~ps. In this case the overlap of the two driving laser pulses is small and the spectra do not show the AT splitting any longer. The simulation in panel (b) is in good agreement with the measurement in panel (a). This proves that the dressed states can only be resolved during the interaction with the first laser pulse. Following the signal intensity from small pulse areas to large ones we see that it forms two maxima. These are Rabi coherence rotations as we have discussed them in Ref.~\cite{wigger2017exp}.
~\\[-7mm]

~\\[-14mm]
\section*{4. Purcell effect}
\begin{figure}[h]
\centering
\includegraphics[width=0.5\columnwidth]{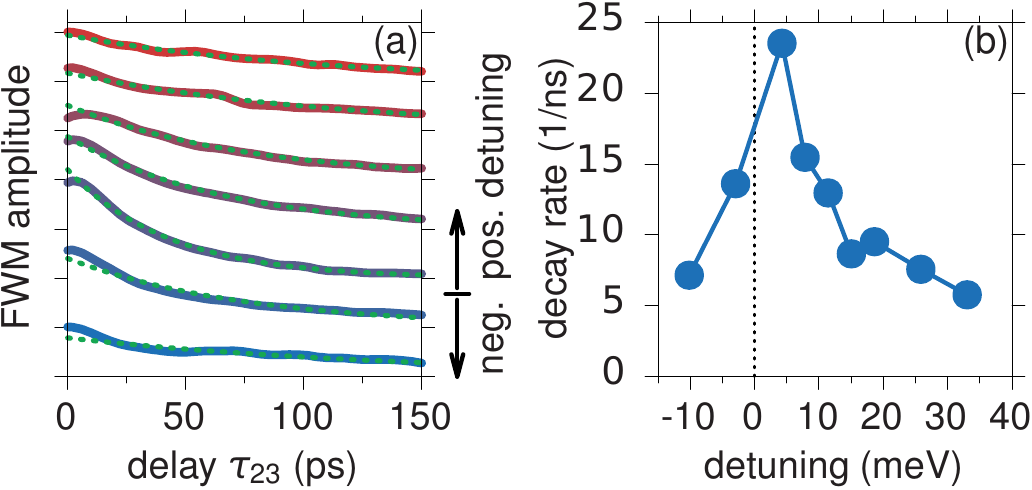}
\caption{Measurement of the Purcell effect. (a) Three-pulse FWM signal dynamics as a function of the delay $\tau_{23}$. The different temperatures, i.e., different detunings, increase from bottom to top. (b) Fitted decay rate as a function of the detuning between the cavity mode and the transition energy.}
\label{fig:purcell_SI}
\end{figure}

By measuring the three-pulse FWM signal in the micropillar-cavity system introduced in the main text, we can probe the radiative lifetime of the exciton state. In the main text we have shown that for small applied pulse areas the dynamics as a function of the delay $\tau_{23}$ is governed by the radiative decay of the excited state population. The signal can be fitted by a single exponential. In Fig.~\ref{fig:purcell_SI}(a) we show the measured FWM signal dynamics as colored solid lines. The corresponding exponential fits are given as green dashed lines. The temperature of the system increases from bottom to top trespassing the crossing in Fig.~\ref{fig:PL_Temp}. Together with the detuning curves in Fig.~\ref{fig:PL_Temp} we can transfer the temperature of the system into a detuning between the cavity mode and the exciton transition. The fitted decay rates from Fig.~\ref{fig:purcell_SI}(a) are plotted against the detuning in panel (b). We find the strongest radiative decay rate in the vicinity of the resonance, i.e., for vanishing detuning. For increasing negative and positive detunings the decay rate decreases rapidly. The reason for this result is the Purcell effect\cite{purcell1946spo,gerard2001ina} which gives rise to a decreasing radiative lifetime when the emitter is coupled stronger to the optical mode due to an increased density of photon states.
~\\[-7mm]

\newpage
\section*{5. Pulse area calibration}
\begin{figure}[h]
\centering
\includegraphics[width=0.6\columnwidth]{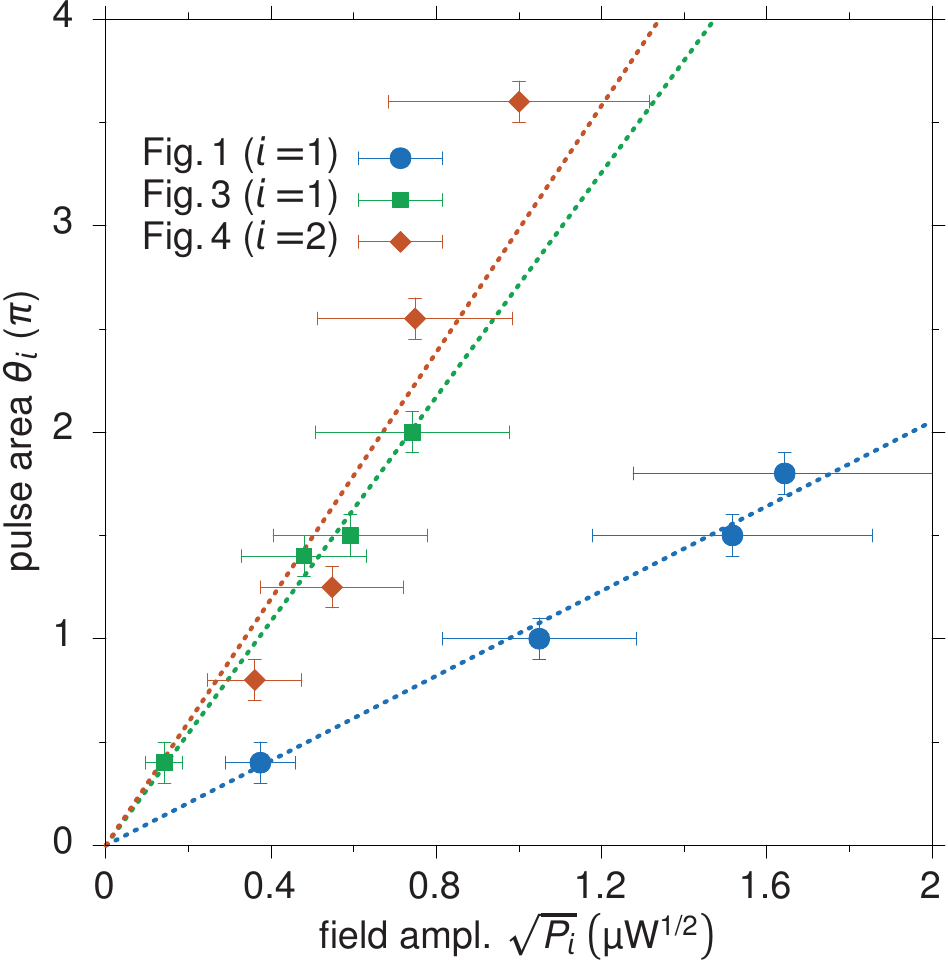}
\caption{Pulse areas calibration. Fitted pulse areas as function of applied field amplitude from the Figures in the main text as given in the plot.}
\label{fig:pulse_areas}
\end{figure}

In the main text we compare the dynamics of measured FWM signals for specific laser pulse powers with simulations, where certain pulse areas are considered to produce the best agreement with the experiment. In Fig.~\ref{fig:pulse_areas} we plot the fitted pulse areas $\theta_i$ as a function of the square-root of the applied laser power $\sqrt{P_i}$, which is proportional to the field amplitude and therefore also to the pulse area in the experiment. Note, that for the results in Fig.~1 in the main text we used another QD sample than for the results in Figs.~4 and 5. For the uncertainties of the applied pulse power we assume 10\% of $P_i$ for the data from the micro-pillar cavity (Figs.~4 and 5) and 5\% of $P_1$ for the planar cavity (Fig.~1). The dotted lines in the plot are linear fits to the data points and serve as a guide to the eye. We find that the dependence between the field amplitudes in the experiment and the pulse areas in the theory follows the expected linear trend. The slope of the Fig.~4 and Fig.~5 data is similar, while one for Fig.~2 is flatter. This is in line with the smaller Q-factor of the planar cavity used for Fig.~2.

\newpage
\section*{6. Complex FWM signal}
\begin{figure}[h!]
\centering
\includegraphics[width=0.6\columnwidth]{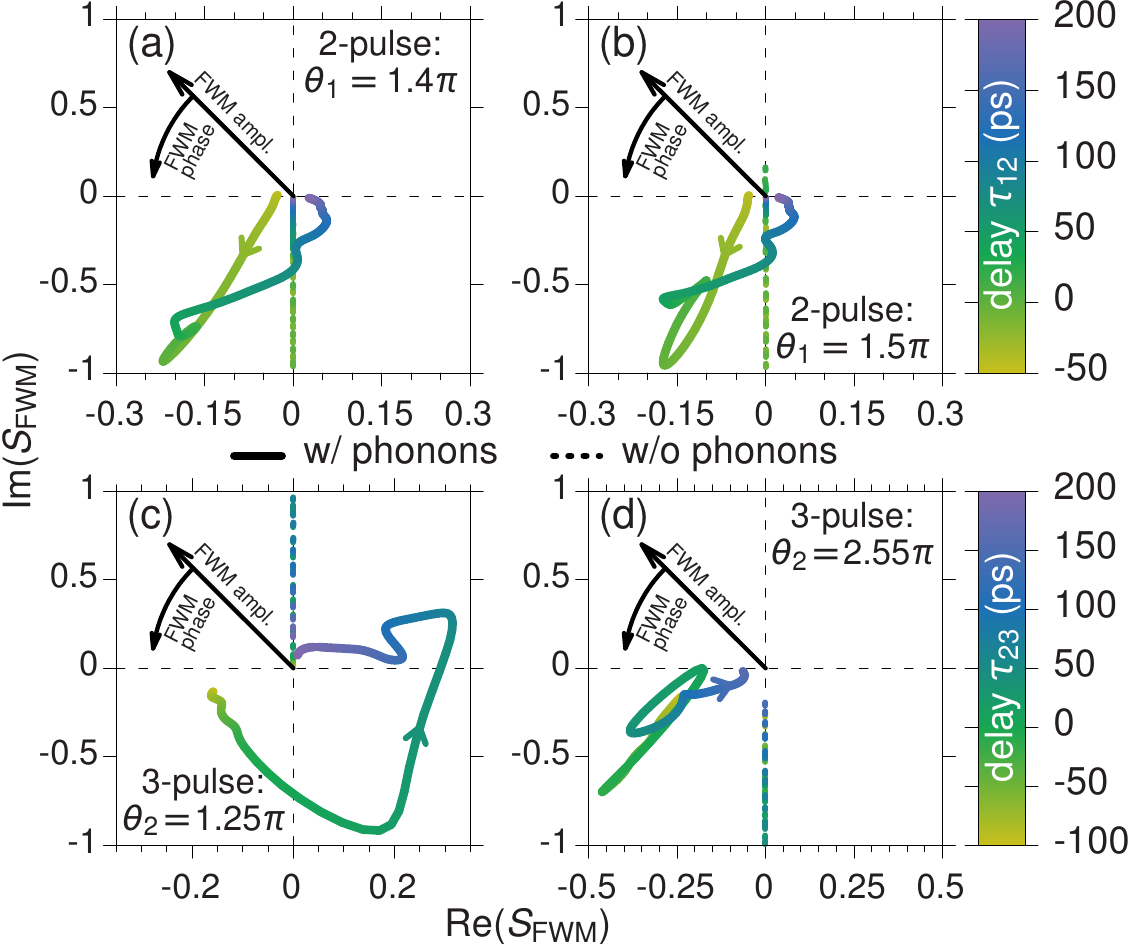}
\caption{ Entire complex FWM signal. Real and imaginary part of the FWM signal $S_{\text{FWM}}$. The delays are color-coded, the dotted lines show simulations without phonon coupling and the solid lines with phonon coupling. (a, b) For two-pulse FWM. The corresponding dynamics of the FWM amplitude is given in Fig.~4(b) in the main text. (c, d) For three-pulse FWM. The corresponding dynamics of the FWM amplitude is given in  Fig.~5(d) in the main text.}
\label{fig:bloch_SI}
\end{figure}

Figure~\ref{fig:bloch_SI} shows the real and imaginary part of the FWM signals with the color coded delays $\tau_{12}$ and $\tau_{23}$. In Fig.~6 in the main text we show the smallest and the largest considered pulse areas in the micropillar-cavity system. Fig.~\ref{fig:bloch_SI} completes the considered pulse areas as it shows the remaining intermediate pulse areas from Figs.~4(b) and 5(d) from the main text. We see that by increasing the pulse areas an increasing number of loops appears in the plots. These loops represent the Rabi oscillations of the Bloch vector during the interaction with the laser pulses.

\end{document}